\documentclass[fleqn,usenatbib]{mnras}

\usepackage[T1]{fontenc}

\DeclareRobustCommand{\VAN}[3]{#2}
\let\VANthebibliography\thebibliography
\def\thebibliography{\DeclareRobustCommand{\VAN}[3]{##3}\VANthebibliography}

\usepackage{graphicx}	
\usepackage{amsmath}	
\usepackage{amssymb}	
\usepackage{lipsum}
\usepackage{xcolor}
\usepackage{nccmath}
\usepackage{comment}
\usepackage{newtxtext,newtxmath}

\title[Radio beam error characterization]{Characterizing Beam Errors for Radio Interferometric Observations of Reionization}

\author[A. Nasirudin et al.]{
Ainulnabilah Nasirudin,$^{1}$\thanks{E-mail: ainulnabilah.nasirudin@sns.it}
David Prelogovic,$^{1}$
Steven G. Murray,$^{2}$
Andrei Mesinger,$^{1}$
and Gianni Bernardi$^{3}$$^{, 4}$$^{, 5}$
\\
$^{1}$Scuola Normale Superiore, Piazza dei Cavalieri 7, I-56126 Pisa, Italy\\
$^{2}$School of Earth and Space Exploration, Arizona State University, Tempe, AZ \\
$^{3}$INAF-Istituto di Radioastronomia, via Gobetti 101, 40129 Bologna, Italy \\
$^{4}$Department of Physics and Electronics, Rhodes University, PO Box 94, Grahamstown, 6140, South Africa\\
$^{5}$South African Radio Observatory (SARAO), 2 Fir Street, Observatory, Cape Town, 7925, South Africa\\
}

\date{Accepted XXX. Received YYY; in original form ZZZ}

\pubyear{2021}

\begin{document}
\label{firstpage}
\pagerange{\pageref{firstpage}--\pageref{lastpage}}
\maketitle

\begin{abstract}
A limiting systematic effect in 21-cm interferometric experiments is the chromaticity due to the coupling between the sky and the instrument. This coupling is sourced by the instrument primary beam; therefore it is important to know the beam to extremely high precision. Here we demonstrate how {\it known} beam uncertainties can be characterized using databases of beam models.  In this introductory work, we focus on beam errors arising from physically offset and/or broken antennas within a station. We use the public code \textsc{OSKAR} to generate an ``ideal'' SKA beam formed from 256 antennas regularly-spaced in a 35-m circle, as well as a large database of ``perturbed'' beams sampling distributions of broken/offset antennas. We decompose the beam errors (``ideal'' minus ``perturbed'') using Principal Component Analysis (PCA) and Kernel PCA (KPCA). 
Using 20 components, we find that PCA/KPCA can reduce the residual 
of the beam in our datasets by $60-90\%$ compared with the assumption of an ideal beam. Using a simulated observation of the cosmic signal plus foregrounds, we find that assuming the ideal beam can result in $1\%$ error in the EoR window and $10\%$ in the wedge of the 2D power spectrum. When PCA/KPCA is used to characterize the beam uncertainties, the error in the power spectrum shrinks to below $0.01\%$ in the EoR window and $\leq1\%$ in the wedge. Our framework can be used to characterize and then marginalize over uncertainties in the beam for robust next-generation 21-cm parameter estimation. 
\end{abstract}




\section{Introduction}

Measuring the Epoch of Reionization (EoR) 21-cm signal is one of the key science goals of current and upcoming low-frequency interferometers such as the Murchison Widefield Array (MWA) \citep{tingay2013murchison,wayth2018}, the Hydrogen Epoch of Reionization Experiment (HERA) \citep{deboer2017hydrogen}, the Low Frequency Array (LOFAR) \citep{vanhaarlem13}, the Giant Metrewave Radio Telescope (GMRT) \citep{swarup1991giant} and the upcoming Square Kilometre Array (SKA) \citep{dewdney2009square,mellema13}. To date, several upper limits of the EoR power spectrum have been published (see e.g. \cite{barry2019improving}, \cite{li2019first}, \cite{trott2020deep}, \cite{mertens2020improved}, and \cite{abdurashidova2021first}). For an actual detection of the EoR signal, an unprecedented level of precision is required. This is because foregrounds and instrumental systematics dominate over the reionization signal by several orders of magnitude. 

Possibly the most complicated and pronounced instrumental systematic comes from the uncertainty in the model of the primary beam \citep{jacobs2017first, line2018situ, Sutinjo2015characterization}. For the wide field-of-view (FoV) instruments common to 21-cm cosmology, this beam must be well-characterized over essentially the entire sky, including in side-lobes close to the horizon. The beam itself is a highly multi-dimensional quantity, changing over direction, frequency, pointing and polarization. Furthermore, accurate measurements in the far-field regime are incredibly difficult; individual elements are far too large to be characterized with anechoic chambers. While novel techniques such as mapping with pulsars \citep{newburgh2014calibrating} and drones \citep{jacobs2017first} show some promise, it is unclear if they will achieve the necessary angular resolution and coverage required. 

In principle, simulations of the beam via electromagnetic modelling are highly accurate and very illuminating; however they can only be computed for an ideal antenna/station and take significant resources. An SKA ``station" simulation is typically performed by combining electromagnetic simulations of isolated antennas with the station layout. There are several ways that such a simulation can be inaccurate, including that it does not take into account full mutual coupling effects, potential in-situ small physical defects of antennas,  potential deviations of precise locations with respect to the planned layout, and the potential for individual antennas to be offline for a particular measurement.


For interferometers like the SKA, the primary beam pattern is dictated by the dipole array structure in each station but there are several sources of deviations from an ideal primary beam. One important source of deviation is the electromagnetic coupling among closely packed elements \citep{fagnoni2021, sutinjo2020design, bolli2021ska1}. In addition, after the antennas are physically  put on site, many small things can happen to them that would make them respond slightly differently from the ideal. For example, although the layout is accurate when the antennas are first deployed, this may not be true after a certain time has passed, since the physical position of the antenna can be displaced or even be offline due to multiple reasons e.g natural events and work maintenance. Indeed, \cite{joseph2020calibration} find that the MWA has a maximum of $12.5\%$ broken antennas per station at any given time during the observation period. 

Accurate models of the beam are incredibly important for 21-cm calibration and power spectrum estimation. Uncertainties in the beam model must be propagated through to parameter estimation. However, this requires defining a suitably compact basis in which to represent the uncertainties, since the beam itself in angular/frequency/time/polarization space is far too complex to propagate. Thus in this paper, we develop a method for describing the residuals of the beam model with respect to an `ideal' simulation using PCA-basis sets for compression. Our general framework can be applied to any beam uncertainty that can be modeled.   As a demonstration here we use simplified, but realistic, beam perturbations due to  two sources of error: antenna displacement within a station, and the possibility that some fraction of antennas are dead. 

The paper is organized as follows. We first describe the beam simulation and the basis set that we use in \S \ref{sec:methods}. In \S \ref{sec:results}, we present the results of our analysis that motivates our choice for the perturbation model and the impacts on the reconstructed beam errors. In \S \ref{sec:impact_PS}, we study the impact of using the ideal beam and the reconstructed beam errors on the power spectrum of a mock sky consisting of the 21-cm signal and foregrounds in an interferometric framework. Finally, we discuss our findings and conclude the paper in \S \ref{sec:conclusion}.

\section{Method for characterizing beam uncertainties}
\label{sec:methods}

In this section, we explain the method used to generate the beam and the basis set used to characterize the beam errors. We generate electromagnetic simulations of an idealised beam, as well as thousands of realizations of non-ideal beams. This training set is used in two PCA-based approaches to identify the primary modes of variation amongst the perturbed beams, and to determine the number of such modes that adequately reconstruct any perturbed beam. The linear coefficients of these modes thus represent a compressed parameter space which can be marginalized in parameter estimation. We first create a database of perturbed beam realizations and sample the errors in \S \ref{sec:database}. We then introduce our PCA and kPCA methods for characterizing the beam errors in \S \ref{sec:basis_set}. 

\subsection{Database of beam models}
\label{sec:database}
We use \textsc{OSKAR}\footnote{https://github.com/OxfordSKA/OSKAR} to simulate the primary beam response of a station based on the antenna layout. The station has 256 antennas positioned within a circle of diameter 35m, motivated by the design of upcoming SKA stations. We define the ideal station as one in which all the antennas are regularly-spaced within the circle, with a spacing of $\sim 1.8$m between each antenna\footnote{The actual SKA stations are expected to have pseudo-randomly distributed antennas, but this assumption does not affect the goal of this paper.}. We generate the beam at 150, 170 and 190 MHz for  a zenith-pointing observation. The exact input parameters we use in \textsc{OSKAR} are presented in Table \ref{tbl:OSKAR_params} in Appendix \ref{app:extra}.

Our perturbation model is based on two scenarios:
\begin{enumerate}
\item broken (i.e. offline) antennas, in which some antennas are excluded from the beam synthesis process. The number of broken antennas in each realization is sampled from a uniform distribution between $N_{\rm broken}=$ 1 -- 12 (corresponding to $\lesssim 5\%$ of the total 256 antennas), with their positions assigned randomly.
\item offset antennas, in which \textit{all} antennas are displaced from the ideal position following a zero-mean normal distribution with $\sigma_x=\sigma_y=3\,$cm.\\
\end{enumerate}

\begin{figure}
\includegraphics[width=\linewidth]{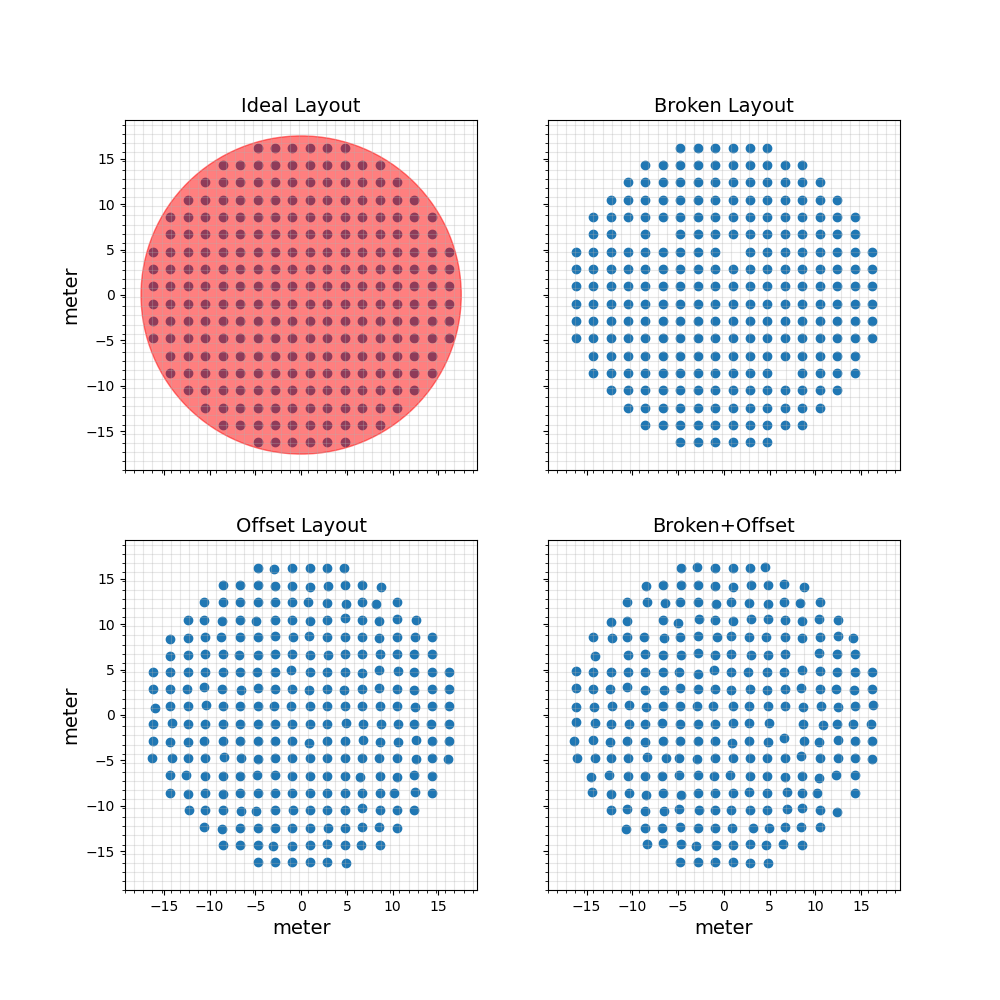}
\caption{\small\label{fig:antenna_layout} The ideal antenna layout (top left panel), along with the same layout but with broken (top right panel), offset (bottom left panel), and both broken+offset antennas (bottom right panel).}
\end{figure}

 We create three separate datasets, each comprised of 10,000 perturbed beam realizations: (i) broken+offset, (ii) broken only, and (iii) offset only. 
 Figure \ref{fig:antenna_layout} shows the ideal antenna layout, along with three examples of the same layout but with broken-only, offset-only, and both broken+offset antennas, respectively.
The 10,000 realizations in each dataset are divided into 7,000 training and 3,000 test realizations.

\begin{figure*}
\includegraphics[width=\linewidth]{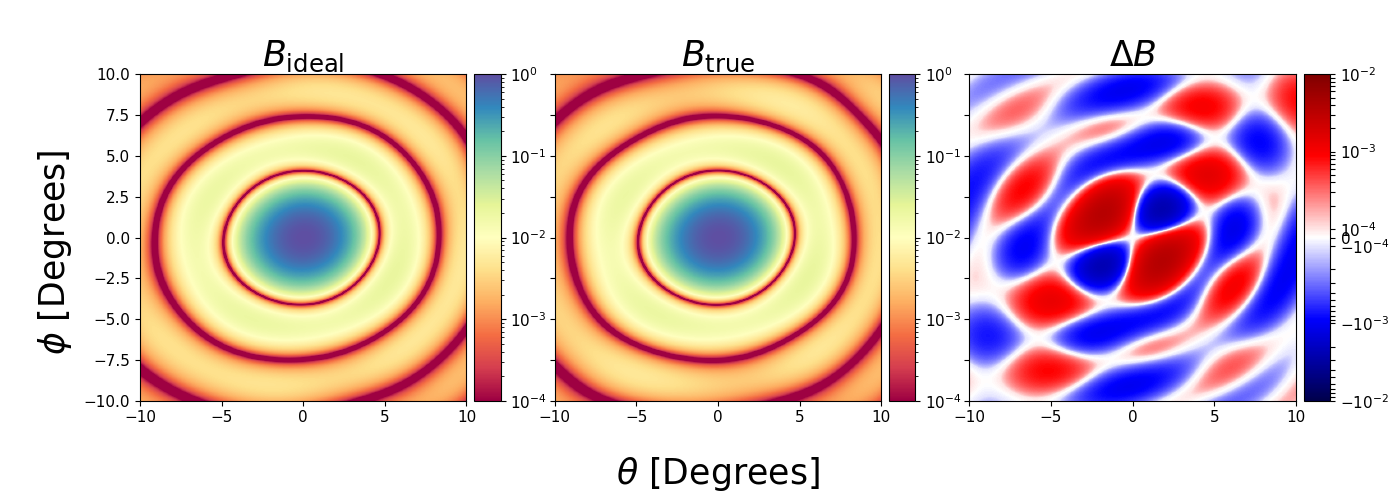}
\caption{\small\label{fig:beam_example} The ideal beam, $B_{\textrm{ideal}}(\nu)$ (left panel), an example realization of $B_{\textrm{true}}(\nu)$ (middle panel) broken+offset data, and the consequent $\Delta B (\nu)$ (right panel) at $\nu =$150 MHz.}

\includegraphics[width=\linewidth]{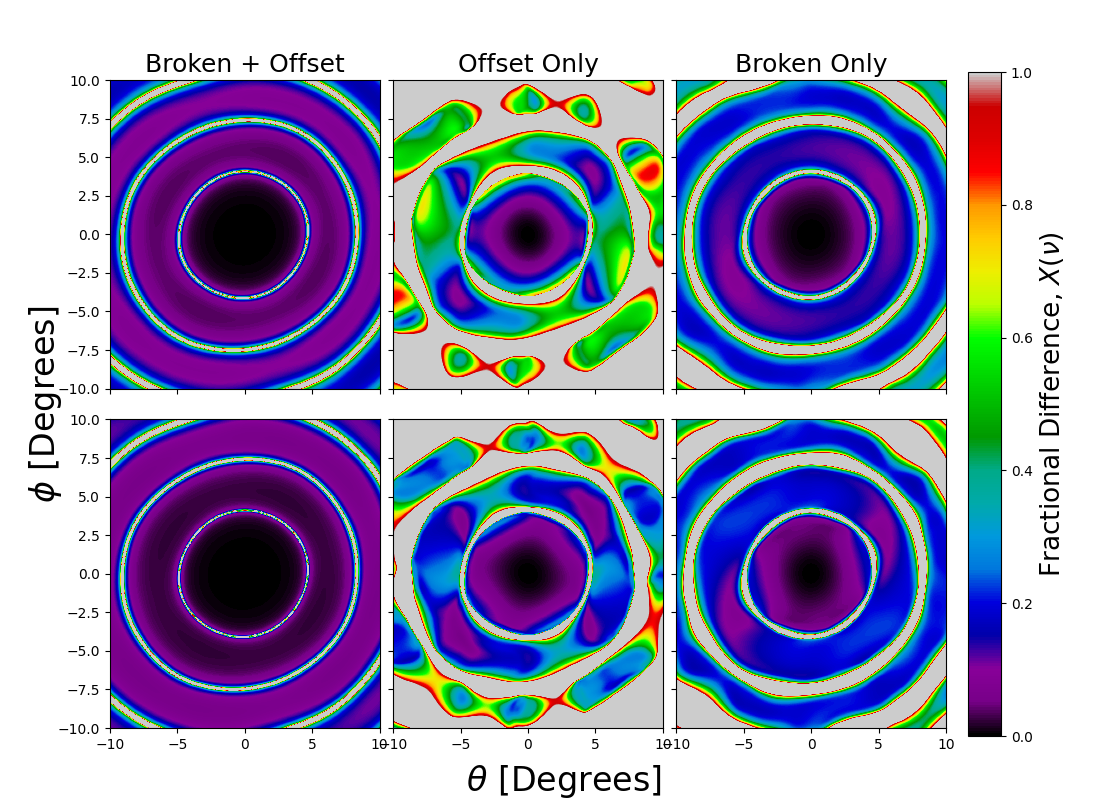}
\caption{\small\label{fig:fracDiff_beam} Mean of the fractional residual, $\mu(X)$ (top panels), and their standard deviations, $\sigma(X)$ (bottom panels), over the 10,000 realizations for the broken+offset (left), offset-only (middle) and broken-only (right) datasets at 150 MHz, where the maximum value for the colour-scale has been limited to 100$\%$.}

\end{figure*}

Throughout, we refer to the simulated power beam of the ``ideal" station as $B_{\rm ideal}(\nu, \theta, \phi)$, where $\nu$ is the frequency, $\theta$ is the zenith angle, and $\phi$ the angle around the zenith pole, and the the simulated power beam of a particular realization of a perturbed station as $B_{\rm true}(\nu, \theta, \phi)$. Instead of modeling the perturbed beams themselves, we model their \textit{residuals}, i.e:
\begin{ceqn}
\begin{equation}
\label{eqn:residual}
 \Delta B( \nu, \theta, \phi) = B_{\textrm{true}}( \nu, \theta, \phi) - B_{\textrm{ideal}}( \nu, \theta, \phi).
\end{equation}
\end{ceqn}
A particular model of the residuals will be represented as $\widehat{\Delta B}$, where we use subscripts to denote the basis used (see next subsection). Note that the residual of residuals is equivalent to the residual in the modeled beam, i.e. $\Delta B - \widehat{\Delta B} \equiv B_{\rm true} - (B_{\rm ideal} + \widehat{\Delta B}) $. For visualization purposes, in Figure \ref{fig:beam_example} we show the ideal beam, $B_{\textrm{ideal}}$ (left panel), an example realization of $B_{\textrm{true}}$ (middle panel) from the broken+offset data, and the corresponding $\Delta B$ (right panel) at $\nu =$150 MHz.

To quantify the beam errors we calculate the fractional residual, $X=|\Delta B|/B_{\rm ideal}$ for each model in the datasets, and then compute the mean and standard deviation of $X$ over all realizations at each frequency, $\nu$. We present the sample mean (top panels) and sample standard deviation (bottom panels) at 150 MHz for the three sets in Figure \ref{fig:fracDiff_beam}. For the broken+offset (left panels) datasets, the mainlobe areas are mostly unaffected by the different antenna configuration as shown by $\sigma(X)$ that is consistent with 0, although the sidelobes can differ by as much as 100$\%$ near the nulls. In contrast, the impact of the antenna offsets (middle panels) is very large, with errors exceeding 100\% around the nulls and in the side-lobes.  Interestingly, the residuals in the broken-only (right panels) dataset are intermediate between the ones in the offset-only and broken+offset.  This could imply that the impact of "breaking" and "offsetting" can partially compensate for one another.
Indeed, one might consider that the positional offsets partially act to ``fill the gaps'' left by the broken antennas, resulting in a net lower error when both effects are at play.


\subsection{Perturbation Basis Set}
\label{sec:basis_set}

We utilize Principal Component Analysis and Kernel Principal Component Analysis to model the beam residuals, $\Delta B$, arising from the different antenna configurations. We describe each in turn.  These are applied to the 7000 samples in each of our three beam error training sets, and then tested on the remaining 3000 samples.

\subsubsection{Principal Component Analysis}
The goal of a traditional Principal Component Analysis (PCA) is to reduce the dimensionality of a dataset by performing a linear change of basis and determining the extent to which each eigenvector captures the variation within the dataset. The most significant basis vectors are termed the ``principal components''. Typically, a dataset $Y= (y_1, y_2, ..., y_{n-1}, y_n)$ is first standardized with respect to its mean and standard deviation,
\begin{ceqn}
\begin{equation}
  Z = \frac{Y - \mu (Y)}{\sigma(Y)},
\end{equation}
\end{ceqn}
and its covariance matrix, $C(Z)$, is computed. The eigenvectors, \textbf{v}, and eigenvalues, $a$, are then calculated following 
\begin{ceqn}
\begin{equation}
  C \textbf{v} = a \textbf{v}.
\end{equation}
\end{ceqn}

Finally, $a$ are arranged in descending order, yielding principal components in order of significance in which a feature vector can be formed with some number of features or components, $N$, and the reconstructed beam, $\widehat{\Delta B}_{N; \rm  PCA}= \sum_i^N a_i \textbf{v}_i$. The principal components have the limitation that they are \textit{linear} transformations of the input dataset; non-linear transformations that require fewer terms to adequately describe the data may exist.\\

\subsubsection{Kernel Principal Component Analysis}
Kernel Principal Component Analysis (KPCA) extends the PCA method via non-linear transformations of the dataset.  The data is first mapped to an arbitrary higher dimension, often referred to as the \textit{feature space}, and then linear PCA is performed on this feature space. The feature space, however, does not need to be explicitly computed. Instead, it is sufficient to compute the kernel,
\begin{ceqn}
\begin{equation}
  K(y_i, y_j) = \psi(y_i)^T \psi(y_j),
\end{equation}
\end{ceqn}
where $\psi(y_i)$ is the non-linear transformation from real to feature space \citep{scholkopf1997kernel}. One downside of KPCA is that a unique, one-to-one inverse relation that transforms $\psi(y_i)$ back to $y_i$ does non exist. However, other methods such as ridge regression \citep{hoerl1970ridge1, hoerl1970ridge2} can be used for this purpose, which is what is being used here. For a simple introduction to KPCA, we refer the reader to Appendix \ref{app:kpca_example}. \\

To model our dataset comprised of $\Delta B(\nu)$, we developed \textsc{SPax}\footnote{https://github.com/dprelogo/SPax
}, an efficient PCA and KPCA code that is GPU and CPU-optimized. The following kernels are available within \textsc{SPax} for KPCA:
\begin{ceqn}
\begin{align}
\label{eqn:kernel}
K(y_i, y_j) &= y_i^T y_j \, \textrm{[linear]}\\
&= \textrm{tanh}\left(\kappa y_i^T y_j\right) \, \textrm{[tanh]}\\
&= \left(\kappa y_i^T y_j + r\right)^d \, \textrm{[polynomial]}\\
&= \exp \left( - \kappa [|y_i + y_j|]^2 \right)\textrm{[radial basis]}\\
&= y_i^T y_j / (y_i^2 \cdot y_j^2) \, \textrm{[cosine]}.
\end{align}
\end{ceqn}
The kernels, $K$, regularization parameter,
and hyper-parameters $\kappa, r,$ and $d$ are flexible;  different kernels (and/or hyper-parameter values) can be used for the transform and inverse transform respectively, in order to improve the fit to the training dataset. We perform hyper-parameter optimization using a simple, coarse grid search, selecting the parameter combination that minimizes the Mean Square Error (MSE) between $\Delta B$ and the reconstructed residual using $N$ features, $\hat{\Delta B}_N$ for all three frequencies. We note that reconstructing the beam error/residual is obtained by finding the best fit set of eigenvalues using the given eigenvector basis. Hyper-parameter calibration is performed separately for all three datasets, and we allow the inverse kernel to be different from the transform kernel.  Our course grid assumes integer values, with $N=10$, $d=2$, and $r=1$ where applicable to reduce computation. The best set of parameters that gives the lowest MSE for each dataset are presented in Table \ref{tbl:kpca_params}, in which the subscript ``inv'' refers to the parameter for the inverse transform.  We highlight that using more sophisticated hyper-parameter Bayesian optimization should yield even better results; we defer this to future work when we apply our method to mock data.


The relative performance of KPCA vs. (linear) PCA can depend strongly on the processes which generate the data itself. In simplest terms, if the data itself is a linear combination of effects, then PCA is optimal. However, 
if the data is inherently a nonlinear transformation from a more compact basis, then KPCA may be better in compressing the information content.
In practice, if the data is inherently most compact in a non-linear basis, we may expect KPCA to outperform (i.e. have a smaller MSE) linear PCA when reconstructing with a `small' number of components. However, linear PCA is guaranteed to achieve perfect reconstruction if using $N \rightarrow N_{\rm dim}$ (i.e. an MSE of close to zero), whereas KPCA is not\footnote{KPCA essentially performs standard linear PCA in a non-linearly transformed space. While it is guaranteed to minimize the MSE in this space, it is not guaranteed to minimize MSE in the space of the data. It is difficult to judge whether this is better or worse without understanding the natural basis of the data.}, and therefore there may be a crossover at some $N$.

\begin{table}
\begin{tabular}{c|c|c|c|}
\cline{2-4}
\multicolumn{1}{l|}{}                    & \textbf{Broken + Offset} & \textbf{Broken Only} & \textbf{Offset Only} \\ \hline
\multicolumn{1}{|c|}{$\kappa$}           & 54                        & 2                   & 79                    \\ \hline
\multicolumn{1}{|c|}{$\kappa_{\rm inv}$} & 1                      & 1                    & 23                   \\ \hline
\multicolumn{1}{|c|}{$K$}                & tanh                     & tanh                  & tanh                 \\ \hline
\multicolumn{1}{|c|}{$K_{\rm inv}$}      & poly                     & rbf                 & poly                  \\ \hline
\end{tabular}
\caption{\label{tbl:kpca_params} The best set of parameters and kernels that gives the lowest MSE for each dataset based on a simple, coarse Monte-Carlo ``grid-search''. The subscript ``inv'' refers to the parameter and kernel for the inverse transform.}
\end{table}

\begin{figure}
\centering
\includegraphics[width=\linewidth]{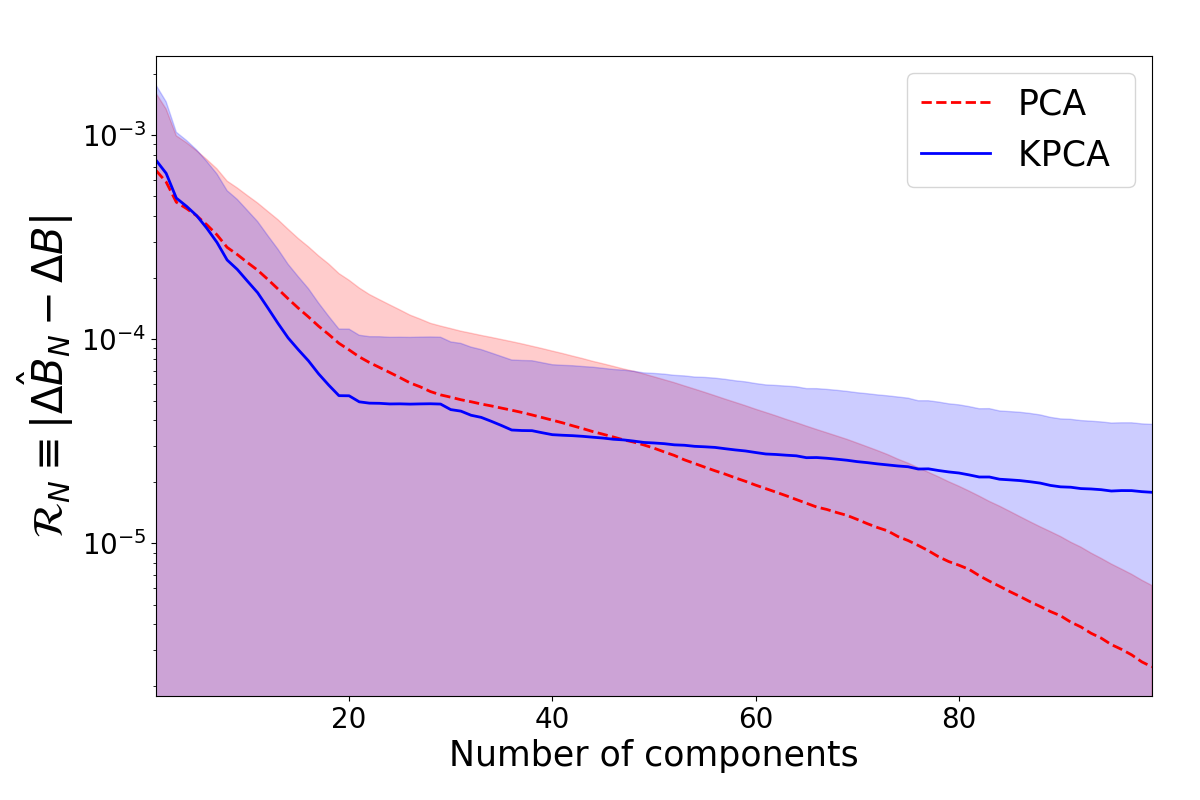}
\caption{\small\label{fig:PCAvsKPCA} The mean (lines) and standard deviation (shaded regions) of $\mathcal{R}_N$ for the broken+offset dataset with varying $N$ (number of components) using PCA (dash red line) and KPCA (solid blue line).}
\end{figure}

\begin{figure*}
\centering
\includegraphics[width=\linewidth]{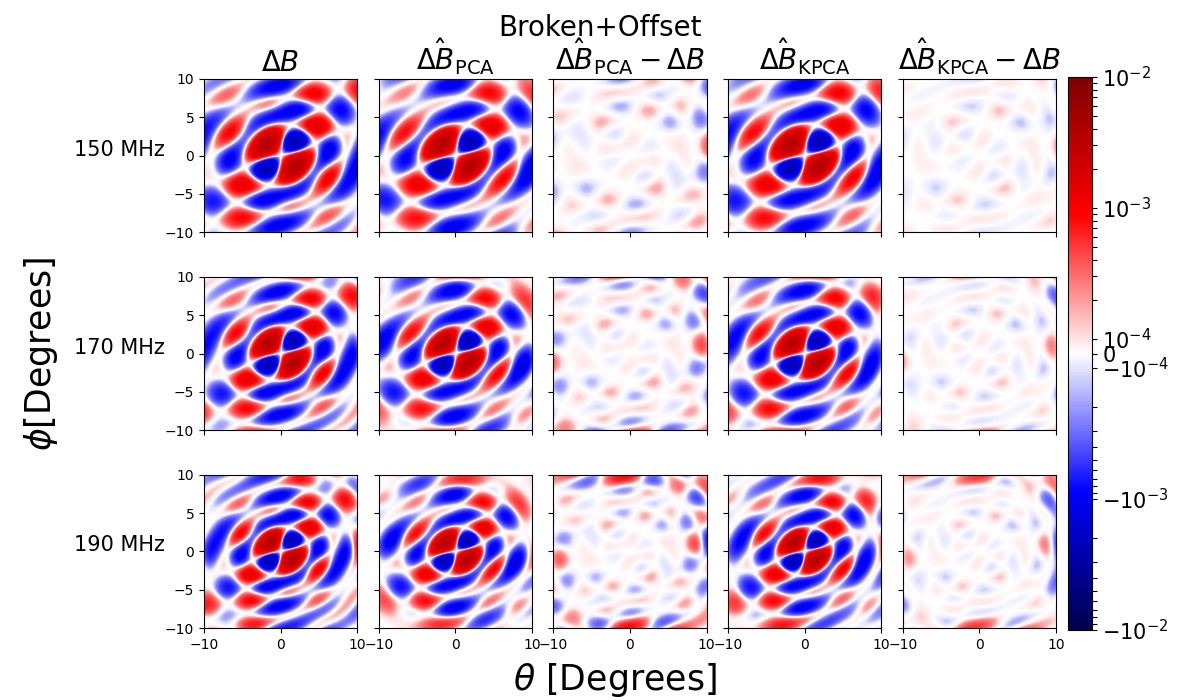}
\caption{\small\label{fig:kpca_beamBrokenOffset} $\Delta B$ (left-most panels), $\hat{\Delta B}$ (second and fourth columns from the left), and $\hat{\Delta B}  - \Delta B$ (middle and right-most columns) from one sample realization of the broken+offset data for all three frequencies.}

\centering
\includegraphics[width=\linewidth]{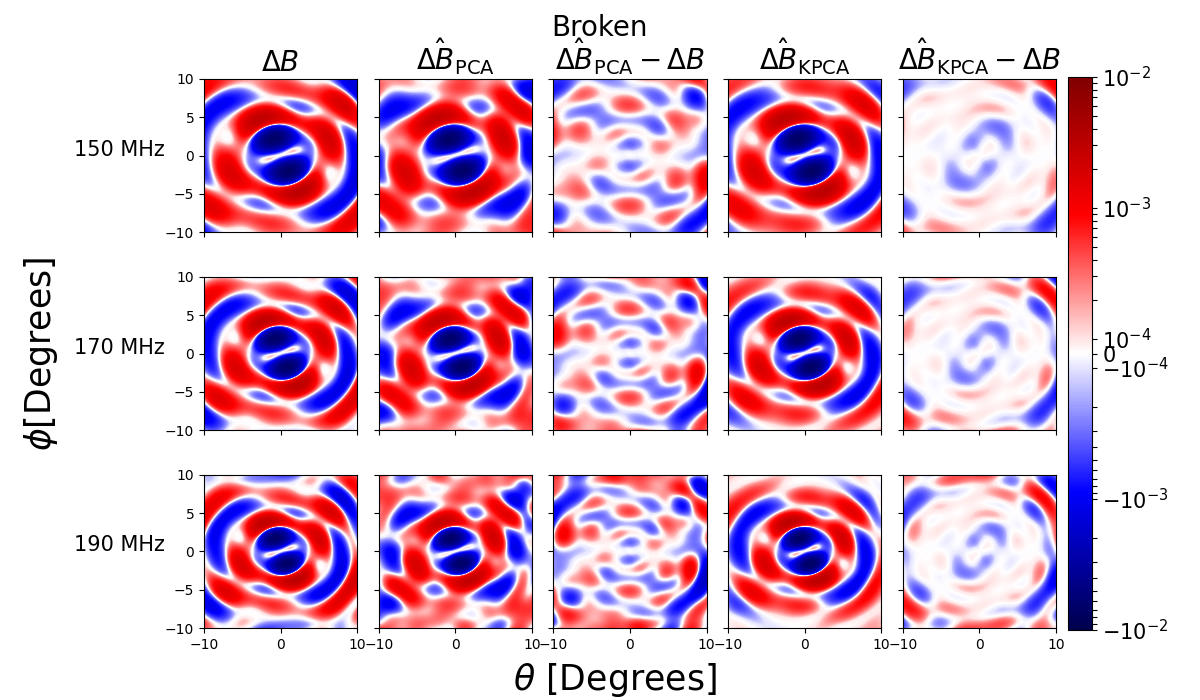}
\caption{\small\label{fig:kpca_beamBrokenOnly} $\Delta B$ (left-most panels), $\hat{\Delta B}$ (second and fourth columns from the left), and $\hat{\Delta B}  - \Delta B$ (middle and right-most columns) from one sample realization of the broken-only set for all three frequencies.}
\end{figure*}

\section{Results: how well is the beam error recovered?}
\label{sec:results}

To decide how many components to include in the reconstruction of the beam across all frequencies, we vary $N$ and evaluate the reconstruction error $\mathcal{R}_N \equiv |\widehat{\Delta B}_N - \Delta B|$. We present the mean, $\langle \mathcal{R}_N \rangle$, and standard deviation, $\sigma(\mathcal{R}_N)$ (lines and shaded regions, respectively) across the 3,000 realizations of broken+offset test data and all three frequencies for PCA (red) and KPCA (blue) in Figure \ref{fig:PCAvsKPCA}. 
Although PCA yields lower values of the mean reconstruction error $\langle \mathcal{R}_N\rangle$ at $N\geq 50$, its decrease with the number of components is slow. In contrast, with KPCA $\langle \mathcal{R}_N\rangle$ decreases rapidly by $N=20$ and then plateaus somewhat as $N$ increases. This relative performance is in qualitative agreement with our expectations from the previous section.

Since we want to model the beam error with the least number of components possible, below we limit ourselves to the first $N=20$ components for both PCA and KPCA. Hereafter, $\widehat{\Delta B}_{N=20}$ is referred to simply as $\widehat{\Delta B}$.

\begin{figure*}

\centering
\includegraphics[width=\linewidth]{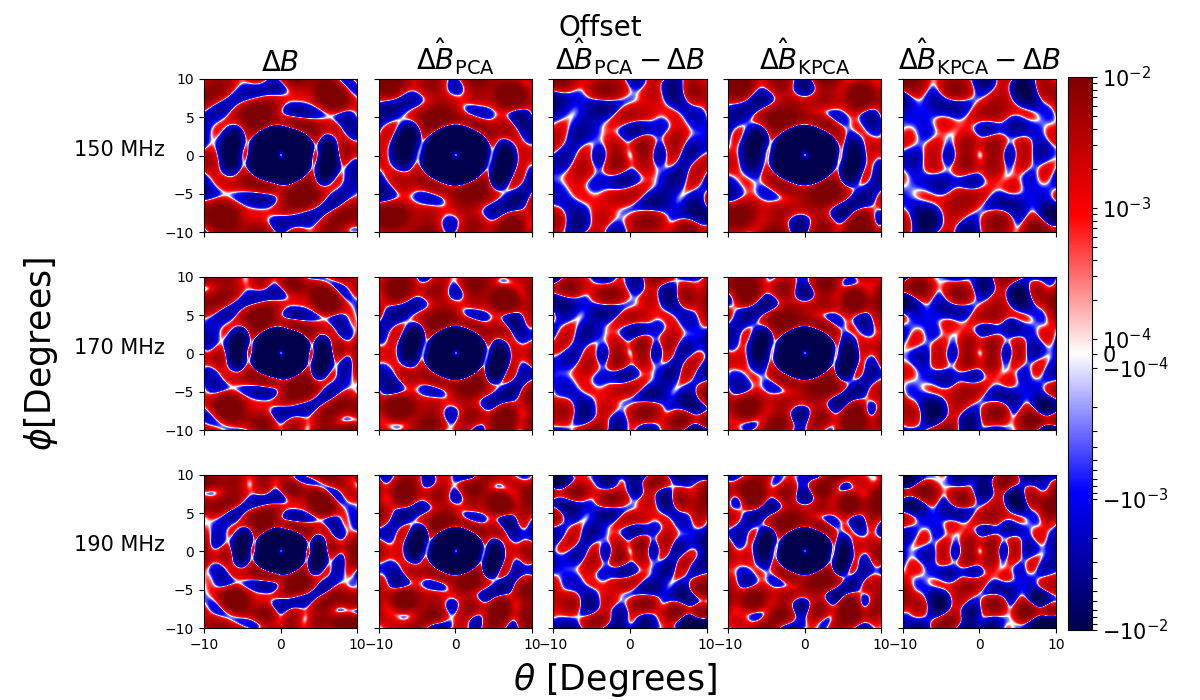}
\caption{\small\label{fig:kpca_beamOffsetOnly} $\Delta B$ (left-most panels), $\hat{\Delta B}$ (second and fourth columns from the left), and $\hat{\Delta B}  - \Delta B$ (middle and right-most columns) from one sample realization of the offset-only set for all three frequencies.}
\end{figure*}

We illustrate the recovery of the beam error in Figures \ref{fig:kpca_beamBrokenOffset} -- \ref{fig:kpca_beamOffsetOnly}, for a randomly-chosen sample from each of our test sets.
From left to right, we show the actual beam error ($\Delta B$), the PCA-reconstructed beam error ($\widehat{\Delta B}_{\rm PCA}$), the difference between the PCA-reconstructed and actual error ($\widehat{\Delta B}_{\rm PCA}  - \Delta B$), the KPCA-reconstructed beam error ($\widehat{\Delta B}_{\rm KPCA}$), and the difference between the KPCA-reconstructed and actual error ($\widehat{\Delta B}_{\rm KPCA}  - \Delta B$).  Rows correspond to our three frequency bins.

With 20 features, both PCA and KPCA can effectively reconstruct $\Delta B$ of our sample from the broken+offset example shown in Fig. \ref{fig:kpca_beamBrokenOffset}, including the frequency evolution of the features, the size of the mainlobe, and the magnitude of the perturbation. All these result in  average difference of $\leq |10^{-4}|$. At 190 MHz, however, the model seems to be slightly less sensitive to structures in the sidelobe region, as is apparent in the middle and right-most panel on the bottom row, especially with PCA. For reference, the distribution of the KPCA eigenvalues (which, as expected, follow a Gaussian distribution), are presented in Figure \ref{fig:distribution_kpca} in the Appendix. 

\begin{table}
\begin{tabular}{l|l|l|l|}
\cline{2-4}
                                                                    & \multicolumn{1}{c|}{\textbf{Broken + Offset}} & \multicolumn{1}{c|}{\textbf{Broken Only}} & \multicolumn{1}{c|}{\textbf{Offset Only}} \\ \hline
\multicolumn{1}{|l|}{$\langle|\Delta B|\rangle$}                               & 3.7$\times 10^{-4}$                          & 8.4$\times 10^{-4}$                      & 5.8$\times 10^{-3}$                      \\ \hline
\multicolumn{1}{|l|}{$\sigma(|\Delta B|)$}                            & 4.5$\times 10^{-4}$                          & 2.3$\times 10^{-3}$                      & 8.5$\times 10^{-3}$                      \\ \hline
\multicolumn{1}{|l|}{$\langle\mathcal{R}_{\textrm{PCA}}\rangle$}     & 4.4 $\times 10^{-5}$                         & 1.9 $\times 10^{-4}$                     & 2.0$\times 10^{-3}$                      \\ \hline
\multicolumn{1}{|l|}{$\sigma(\mathcal{R}_{\textrm{PCA}})$}  & 5.7 $\times 10^{-5}$                         & 1.9 $\times 10^{-4}$                     & 1.8 $\times 10^{-3}$                     \\ \hline
\multicolumn{1}{|l|}{$\langle\mathcal{R}_{\textrm{ KPCA}}\rangle$}   & 3.9$\times 10^{-5}$                          & 1.7$\times 10^{-4}$                      & 2.2$\times 10^{-3}$                      \\ \hline
\multicolumn{1}{|l|}{$\sigma(\mathcal{R}_{\textrm{KPCA}})$} & 4.9$\times 10^{-5}$                          & 3.9$\times 10^{-4}$                      & 2.5$\times 10^{-3}$                      \\ \hline
\end{tabular}
\caption{\label{tbl:comparison_residual} The MSE and standard deviation of MSE of $\Delta B$ and |$\hat{\Delta B}  - \Delta B$| with PCA and KPCA for the three test datasets.}
\end{table}

For the broken-only sample in Fig. \ref{fig:kpca_beamBrokenOnly}, both PCA and KPCA are able to capture the overall details of $\Delta B$, including the evolution of the features and the size of the mainlobe, as shown in the second column from the left and fourth column from left in Figure \ref{fig:kpca_beamBrokenOnly}. However, the error in the reconstruction can be up to an order of magnitude higher with PCA and there are more small scale features compared to the reconstruction with KPCA.

For the offset-only example shown in Figure \ref{fig:kpca_beamOffsetOnly}, both PCA and KPCA perform worse than seen in the previous two examples. The reconstructions (second and fourth columns from left) somewhat resemble the large-scale structures of $\Delta B$, but instead of having two large ``half-ring'' structures in the sidelobe, both PCA and KPCA model them as multiple radial features. Moreover,  the reconstructed error can be up to two orders of magnitude higher than in the previous examples, as is evident in the third and right-most columns.

To summarise, we present the mean and standard deviation of $|\Delta B|$ and $\mathcal{R}_{20}$ over the 3,000 test realizations and all three frequencies for the three test datasets with PCA and KPCA in Table \ref{tbl:comparison_residual}. Using PCA/KPCA, there is up to a factor of ~10 reduction in beam error compared with the assumption of an ideal beam.

\section{Impact of Beam Reconstruction on the Power Spectrum}
\label{sec:impact_PS}

Although we have established that PCA and KPCA do a good job in capturing the beam error from our datasets, the reconstruction is not perfect.
Hence in this section, we investigate the impact of these residual errors on the recovery of the power spectrum, using a realistic sky composed of the EoR signal and point-source foregrounds.

\subsection{Foreground Model}
\label{sec:fg_model}
Following \cite{nasirudin2020impact}, we simulate extra-galactic point-source foregrounds with a flux-density source count distribution with the power-law relation
\begin{ceqn}
\begin{equation}
\label{eqn:dn_ds}
\frac{dN}{dS} (S, \nu)= \alpha S_{\nu}^{-\beta} \left( \frac{\nu}{\nu_0}\right)^{-\gamma \beta} [{\rm Jy}^{-1} {\rm sr}^{-1}],
\end{equation}
\end{ceqn}
where $dN/dS$ is the source spatial density per unit flux density, $S_{\nu}$ is the flux at a specific frequency $\nu$, $\beta$ is the slope of the source-count function, and $\gamma$ is the mean spectral-index of point sources. Based on an observational result from \citet{intema2011deep}, we set $\alpha$ = 4100 Jy$^{-1}$ sr$^{-1}$, $\beta$ = 1.59, and $\gamma = 0.8$ at $\nu_0$ = 150 MHz .
Having drawn source fluxes from the above distribution, we situate them uniformly randomly across the sky. 
We sample the point sources between $S_{\rm min}=50$mJy and $S_{\rm max}=50\mu$
Jy\footnote{\cite{choudhuri2021MNRAS.506.2066C} found that the brightest sources to be particularly important in the presence of beam variations/non-redundancy, but because we assume that brighter sources have been perfectly peeled from the observation, hence we only model those below the peeling threshold.} at 150 MHz. The observation consists of 128 linearly-spaced frequency channels between 150 to 165 MHz.


\subsection{Reionization Model}
\label{sec:eor_model}

The differential brightness temperature, $\delta T_{B}$, during the EoR can be approximated as
\begin{align}
\label{eqn:Tb}
\delta T_{b} (z) \approx & 27 x_{\rm HI} (1+\delta_{\rm nl}) \left(\frac{H(z)}{\mathrm{d}v/\mathrm{d}r +H(z)}\right)\left(1 -\frac{T_\gamma}{T_s}\right) \nonumber \\
& \times \left(\frac{1+z}{10}\frac{0.15}{\Omega_m h^2}\right)^{\frac{1}{2}}\left(\frac{\Omega_b h^2}{0.023}\right)\ [{\rm mK} ],
\end{align}
where $x_\mathrm{HI}$ is the neutral fraction, $\delta_\mathrm{nl}$ is the evolved Eulerian overdensity, $H$ is the evolving Hubble constant, d$v$/d$r$ is the gradient of the line-of-sight velocity component, $T_\gamma$ is the temperature of the CMB, $T_\mathrm{s}$ is the spin temperature of neutral hydrogen (HI), $z$ is the redshift, $\Omega_m$ is the dimensionless matter density parameter, $\Omega_b$ is the dimensionless baryonic density parameter and $h$ is the normalized Hubble constant \citep{furlanetto2006cosmology}.

We use the efficient semi-numerical EoR modelling tool, \textsc{21cmFASTv3} \citep{mesinger201121cmfast,park2019inferring,murray202021cmfast}, to generate the lightcone of $\delta T_b$ during the EoR. 
For a detailed description of the code and astrophysical model, we refer readers to \cite{mesinger201121cmfast}, \cite{park2019inferring} and \cite{murray202021cmfast}. In our research, we use the default parameter values of \textsc{21cmFASTv3}, which are shown to reproduce current high-$z$ observations \citep{park2019inferring} and simulate the lightcone of a $512$ Mpc$/h$ box. This choice of parameters corresponds to a neutral fraction of 0.5 at $z \sim 6.5$. Because the lightcone covers only $\sim 3.3^\circ$ at 150 MHz, we tile it across the 20$^\circ$ mock sky and coarsen the grids to match with the resolution of the beam. 


\subsection{Interferometric Framework}
\label{sec:delay_ps}

At wavelength, $\lambda$, the baseline displacement, \textbf{u} $=(u,v)$, is defined as $\mathbf{u} = \mathbf{x}/\lambda$, where \textbf{x} is the physical displacement between the stations in meters. The sky coordinate, $\mathbf{l}$, is defined as $\mathbf{l} = (l, m) = (\sin \theta \cos \phi, \sin \theta \sin \phi$).

\begin{figure*}
\centering
\includegraphics[width=\linewidth]{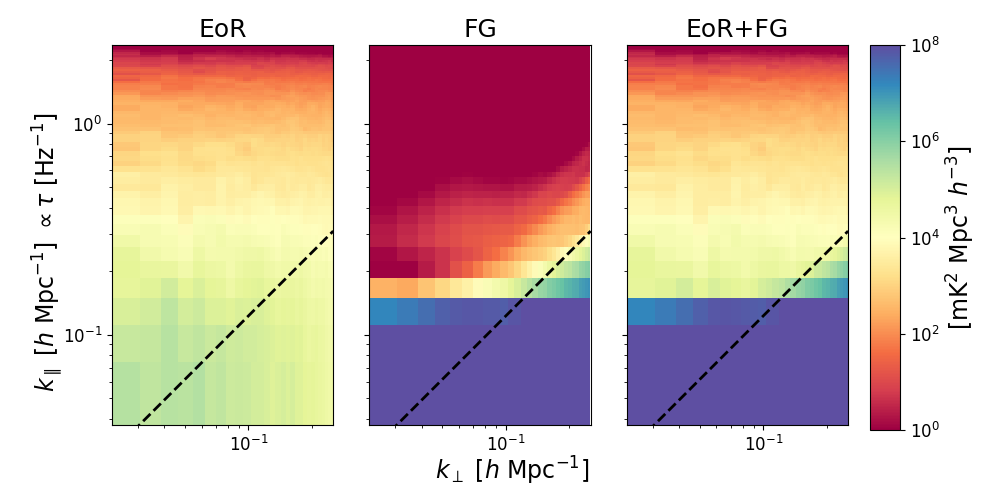}
\caption{\label{fig:diff_ps} From left to right, the PS of cosmic signal, foregrounds, and both cosmic signal and foregrounds respectively, for a mock sky convolved with $B_{\rm true}$. The black dash line shows the extent of the wedge which is calculated following Equation \ref{eqn:wedge_foreground}.}

\centering
\includegraphics[width=\linewidth]{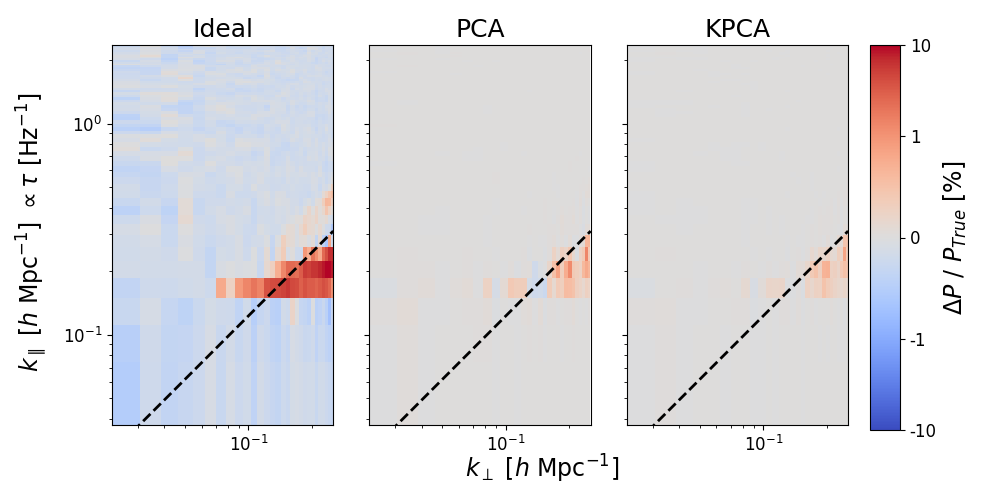}
\caption{\label{fig:frac_ps} From left to right, the fractional difference between the PS of both cosmic signal and foregrounds convolved with $B_{\rm ideal}$, $\hat{B}_{\rm PCA}$, and $\hat{B}_{\rm KPCA}$ with respect to the PS of the same sky convolved with $B_{\rm true}$.}
\end{figure*}

Using the flat-sky approximation, the visibility at frequency $\nu$, $V(\textbf{u}_j,\nu)$, for each baseline $j$ is defined as

\begin{ceqn}
 \begin{equation}
 \label{eqn:FT_foreground}
 V(\textbf{u}_j,\nu) = \int S(\textbf{l}, \nu) B (\textbf{l}, \nu) \exp(-2\pi i \textbf{u}_j\cdot \textbf{l}) d\textbf{l} \ \ \ \mathrm{ [Jy]},
 \end{equation}
\end{ceqn}
where $S(\textbf{l}, \nu)$ and $B(\textbf{l}, \nu)$ are the flux density of each point-source and the beam attenuation at \textbf{l} and $\nu$. The observed interferometric visibility is identical to the Fourier transform of the product of signal and the beam model under the flat-sky approximation. Here, we assume all stations have the exact same layout hence the same beam.

For computation purposes, we Fast Fourier Transform over the 2-D image to a regular-spaced 2D grid $\mathbf{u}_k$, and then interpolate $V(\textbf{u}_k,\nu)$ from the regular 2D grid to the baselines $\textbf{u}_j$. We then apply a Blackman-Harris taper $H(\nu)$ over the frequency axis, and calculate the delay transform (i.e. Fourier transform of un-gridded visibilities along the frequency axis),
\begin{ceqn}
\begin{equation}
\label{eqn:delay_transform}
 V(\textbf{u}_j,\tau) = \int V(\textbf{u}_j, \nu) H(\nu) \exp(-2\pi i \tau \cdot \nu) d\nu \ \ \ [ { \textrm{ Jy Hz}}].
\end{equation}
\end{ceqn}
The delay power spectrum is then calculated by cylindrically averaging the power of the visibilities within radial bin \textbf{r} $=\sqrt{u^2 + v^2}$, which is proportional to the angular mode, $k_\perp$. We approximate the delay power spectrum as the power spectrum, in which $\tau$ is proportional to the line-of-sight mode, $k_\parallel$. The conversion of the power spectrum, \textbf{r} and $\tau$ to cosmological units are outlined in Appendix \ref{apdx:conversion}. For reference, we calculate the wedge region given by

\begin{ceqn}
 \begin{equation}
 \label{eqn:wedge_foreground}
 k_{\parallel} \leq k_\perp \frac{\textrm{sin}(\theta_{\rm FoV})E(z) \int_0^z dz'/E(z')}{(1+z)} \ \ \ [h \ \mathrm{ Mpc}^{-1}],
 \end{equation}
\end{ceqn}
where $\theta_{\rm FoV}$ is the angular radius of the FoV \citep{thyagarajan13,dillon2014overcoming}.

\subsection{Impact of Different Beams on the Power Spectrum}

To understand the impact of beam errors on the recovery of the cylindrical power spectrum (hereafter PS), we convolve our sky described in \S \ref{sec:fg_model} and \ref{sec:eor_model} with the sample realization of $B_{\rm true}$ shown in Figure \ref{fig:kpca_beamBrokenOffset}. Following the steps outlined in \S \ref{sec:delay_ps}, we then simulate the effects of interferometric observation and calculate the respective PS. From left to right, the panels in Figure \ref{fig:diff_ps} show the PS of a sky consisting of the cosmic signal, foregrounds, and both cosmic signal plus foregrounds, respectively, that has been convolved with $B_{\rm true}$. In the foreground PS (middle panel), the well known key features shown in e.g. \cite{dillon2014overcoming} and \cite{barry16} are clearly visible, mainly the foreground dominated region in dark blue where $k_\parallel \leq 0.11 \, h$ Mpc$^{-1}$, the yellow-blue region of the wedge at $k_\perp \geq 0.1 \, h$ Mpc$^{-1}$, and the mostly red EoR window.

Finally in Figure \ref{fig:frac_ps} we present the fractional error in the PS of the beam-convolved total signal (EoR + FG) with respect to the "true" PS from the third panel of Figure \ref{fig:diff_ps}, i.e. $P (B * [{\rm FG+EoR}])/ P(B_{\rm true}*[{\rm FG + EoR}])$ for $B=[B_{\rm ideal}, \widehat{B}_{\rm PCA}, \widehat{B}_{\rm KPCA}]$ (left to right panels respectively). 
From the left panel we see that not accounting for beam errors mis-estimates the power spectrum throughout $k$-space, with errors peaking at $\sim10$\% in the wedge region.  Instead, modeling  $\widehat{\Delta B}$ using either PCA or KPCA reduces the error in the recovered power spectrum by over a factor of a hundred in the EoR window and a factor of ten in the wedge (compare middle and right to the left panel).

Because the visibilities in the wedge are highly correlated, any deficit or surplus of beam attenuation with respect to $B_{\rm true}$ is reflected in the entire wedge region. Indeed, as expected, beam errors affect foregrounds more than the cosmic signal, even in the PS space. 

Finally, we stress that this exercise is highly idealized, providing only a mimimum estimate of the PS recovery error.  In practice, we will not be able to fit the PCA and KPCA coefficients to the true beam directly, as we have done here.  Instead, eigenvalues would need to be co-varied when performing calibration and inference.  We defer this analysis to future work.

\section{Conclusions}
\label{sec:conclusion}
 
Some of the most important systematics in radio interferometry arise from imperfect knowledge of the telescope beam. In this work we demonstrate an empirical approach to characterizing known sources of beam errors.  Focusing on offline and offset antennas for an SKA-like beam, we generate thousands of realizations of beam errors.  We use these realizations to define a beam error basis using PCA and KPCA.

We demonstrate that both PCA and KPCA perform well in recovering beam errors from offline and offset antennae.  Compared with assuming an "ideal" beam, using the top 20 components in either basis can reduce the MSE by $\sim$tens--100\% over our test datasets.

We demonstrate how this beam error characterization translates to improved power spectrum recovery.  We generate a mock sky comprised of point source foregrounds and the cosmic signal, and recover the cylindrical power spectra assuming different beam models.  For a random realization of beam error, we find that assuming the "ideal" beam results in PS errors that peak at $\sim10$\% around the wedge region. Instead if either PCA or KPCA is used to characterize the true beam with 20 components, the PS error is reduced by a factor of $\sim$10--100 throughout $k$-space.

We stress that we did not include additional errors from, e.g. calibration, in this work. We expect that fractional errors in sky-based calibration will be much more sensitive to errors in the assumed beam model, and these are further squared when propagated to power spectrum space. Depending on the spectral structure of these calibration errors, inaccuracies as small as $10^{-5}$ can be crippling to a power spectrum estimation \citep{barry16}.  Therefore, we expect improved beam characterization to be even more important when calibration is also included; we defer this to future work.

Our general framework of using an empirical basis to characterize systematics should prove useful for an end-to-end inference pipeline for 21-cm interferometry.  The principal eigenvectors from PCA and KPCA can provide an optimal basis for systematics, with the corresponding eigenvalues being co-varied together with cosmological parameters when performing Bayesian inference. We will demonstrate this in a follow-up work.

\section*{Acknowledgements}
We thank P. Bull for helpful comments on a draft version of this work.  This work was supported by the European Research Council (ERC) under the European Union’s Horizon 2020 research and innovation programme (grant agreement No 638809 – AIDA – PI: Mesinger). The results presented here reflect the authors’ views; the ERC is not responsible for their use. We gratefully acknowledge computational resources of the Center for High Performance Computing (CHPC) at Scuola Normale Superiore (SNS).
\section*{Data Availability}
The data from this study will be shared on reasonable request to the corresponding author.
 



\bibliographystyle{mnras}
\bibliography{references} 




\appendix

\section{Unit Conversion}\label{apdx:conversion}
The conversion of $\delta T_B$ to $S(\nu)$ (and vice versa) follows the Rayleigh-Jeans law,
\begin{ceqn}
\begin{equation}
\label{eqn:mK_to_Jy}
S(\nu) = \left(\frac{2k_B \nu^2 \delta T_B}{c^2}\right) \times 10^{26}  \ \ \ [{\rm Jy/sr}],
\end{equation}
\end{ceqn}
where $k_B$ is the Boltzmann constant.

Under the assumption that $\tau$ is equivalent to the Fourier counterpart of the line-of-sight mode, $\eta$, both $k_\perp$ and $k_\parallel$ are converted from \textbf{r} and $\tau$ in Fourier dimensions following 

\begin{ceqn}
\begin{equation}
\label{eqn:kperp_modes}
k_\perp = \frac{2 \pi |\textbf{r}|}{D_M(z)}\ \ \ [ {\rm Mpc^{-1} h}],
\end{equation}
\end{ceqn}

and

\begin{ceqn}
\begin{equation}
\label{eqn:kpar_modes}
k_\parallel = \frac{2 \pi H_0 f_{21} E(z) }{c(1+z)^2} \tau \ \ \ [ {\rm Mpc^{-1} h}]
\end{equation}
\end{ceqn}
from \cite{morales2010reionization}. Here, $z$ is the observation redshift, $D_M(z)$ is the transverse comoving distance, $H_0$ is the Hubble constant, $f_{21}$ is the rest frequency of the 21\,cm hydrogen hyperfine transition and $E(z)$ is defined as

\begin{ceqn}
 \begin{equation}
\label{eqn:Ez_cosmology}
E(z) = \sqrt[]{\Omega_m(1+z)^3+\Omega_k(1+z)^2+\Omega_\Lambda},
\end{equation}
\end{ceqn}

where $\Omega_\Lambda$, and $\Omega_k$ are the dimensionless density parameters for dark energy and the curvature of space.

\section{Extra Materials}
\label{app:extra}

\begin{figure}
\centering
\includegraphics[width=\linewidth]{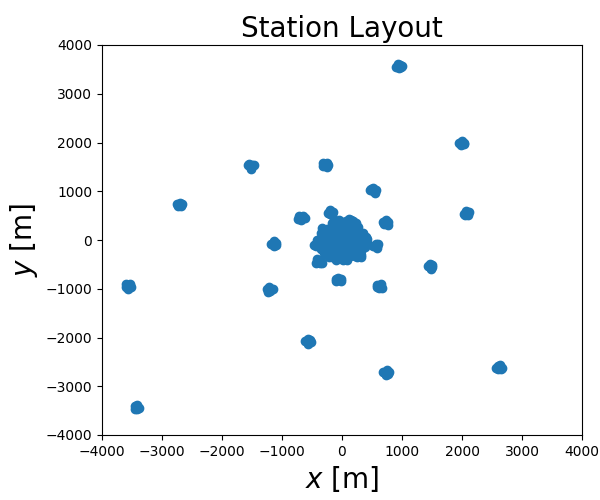}
\caption{\small\label{fig:stations} The SKA-like station layout.}
\end{figure}

\begin{figure}
\centering
\includegraphics[width=\linewidth]{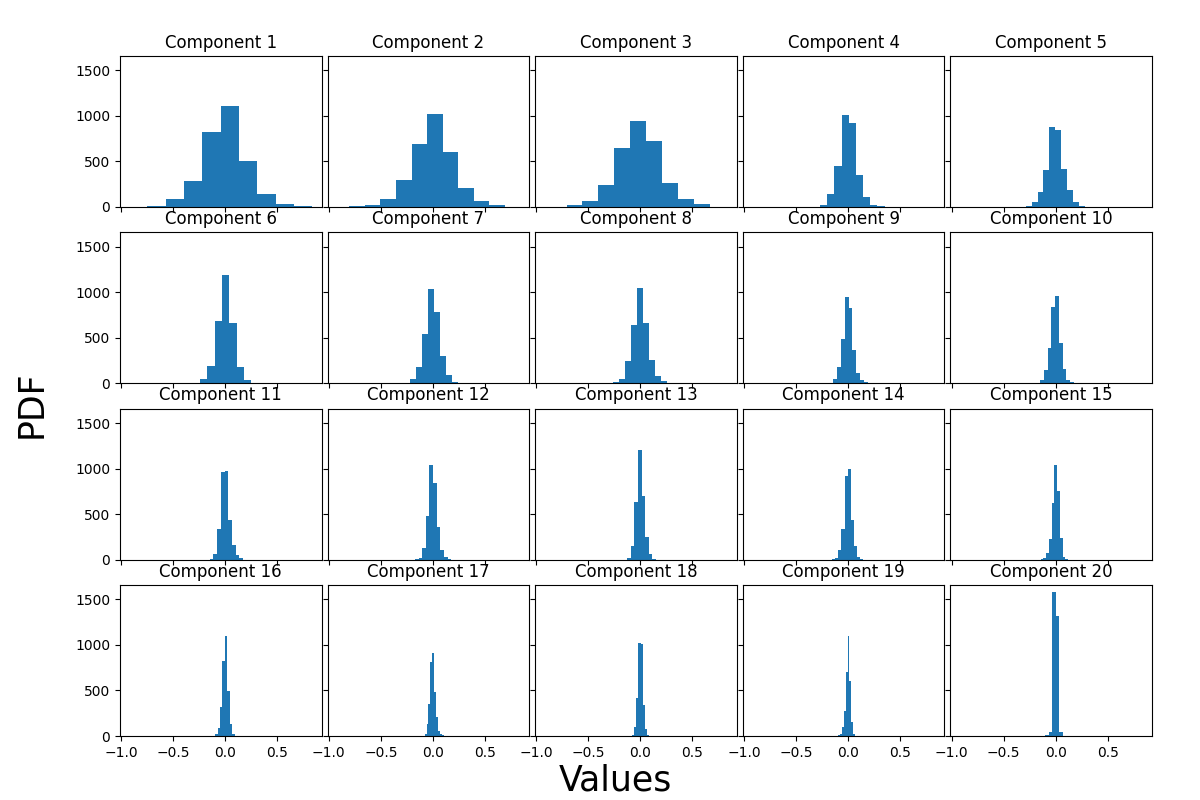}
\caption{\small\label{fig:distribution_kpca} The Probability Density Functions (PDFs) of the first 20 components in the higher dimension space from the KPCA, ordered from largest (top left panel) to smallest (bottom right panel) variance.}
\end{figure}

In this section, we present some extra materials concerning the research for interested readers. The parameter input used for OSKAR is presented in Table \ref{tbl:OSKAR_params} and the SKA-like station layout is shown in Figure \ref{fig:stations}. In addition, Figure \ref{fig:distribution_kpca} shows the Probability Density Functions (PDFs) of the first 20 components in the higher dimension space from the KPCA, ordered from largest (top left panel) to smallest (bottom right panel) variance.

\begin{table}
\begin{tabular}{|c|c|}
\hline
\textbf{Parameter}        & \textbf{Values} \\ \hline
FoV ($^\circ$)          & 20       \\ \hline
RA of Observation ($^\circ$)   & 0        \\ \hline
Dec of Observation ($^\circ$)   & -27       \\ \hline
Latitude of Telescope ($^\circ$) & -27       \\ \hline
Longitude of Telescope ($^\circ$) & 117       \\ \hline
Observation Time (UTC)      & 17:00:00    \\ \hline
Observation Date         & 1 August 2020  \\ \hline
Frequencies (MHz)         & [150, 170, 190] \\ \hline
\end{tabular}
\caption{\label{tbl:OSKAR_params} The \textsc{OSKAR} parameter input used in this research.}
\end{table}

\section{Kernel PCA example} \label{app:kpca_example}
The purpose of this simple example is to qualitatively describe the main ingredients of the KPCA algorithm - in particular: data space - $y$, feature space - $\psi(y)$, and two kernels ($\kappa$ and $\widetilde{\kappa}$) defining mappings from one to another. Moreover, we would like to show very different roles the two kernels have in the process.

\begin{figure}
     \centering
     \includegraphics[width=0.9\linewidth]{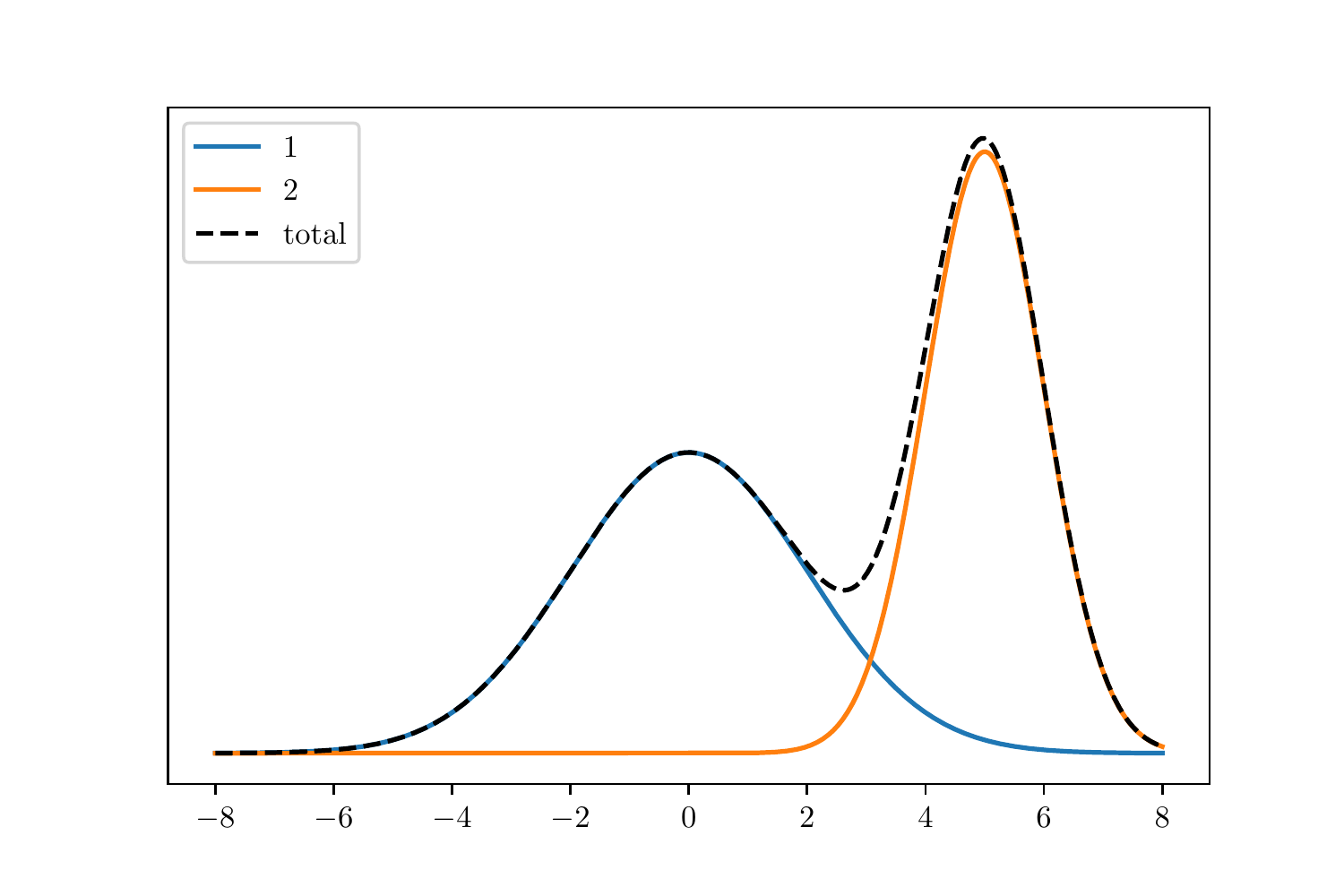}
     \caption{Initial distribution on which we would like to run a KPCA algorithm.}
     \label{fig:init_distribution}
\end{figure}

In Figure \ref{fig:init_distribution} we show a bi-modal distribution (with the two modes labeled '1' and '2').  The horizontal axis represents data space.
In this simple example, we would like to use KPCA to accentuate the bi-modality of this distribution.
We first pull samples from the distribution and use the kernel $\kappa(y_i, y_j) = \psi(y_i) \cdot \psi(y_j)$ as a part of the KPCA algorithm. Here $\psi(y)$ is in feature space, which is not directly accessible and can be infinite dimensional.  Performing KPCA in this space and selecting the first $N$ components amounts to selecting basis vectors in the feature space following the largest variance of the samples. After fitting the data, this subset of a feature space $\boldsymbol{\psi}_N$ is accessible and the mapping $\boldsymbol{\psi}_N (y)$ is known.

\subsection{Mapping to and from feature space}

\begin{figure}
     \centering
     \includegraphics[width=0.9\linewidth]{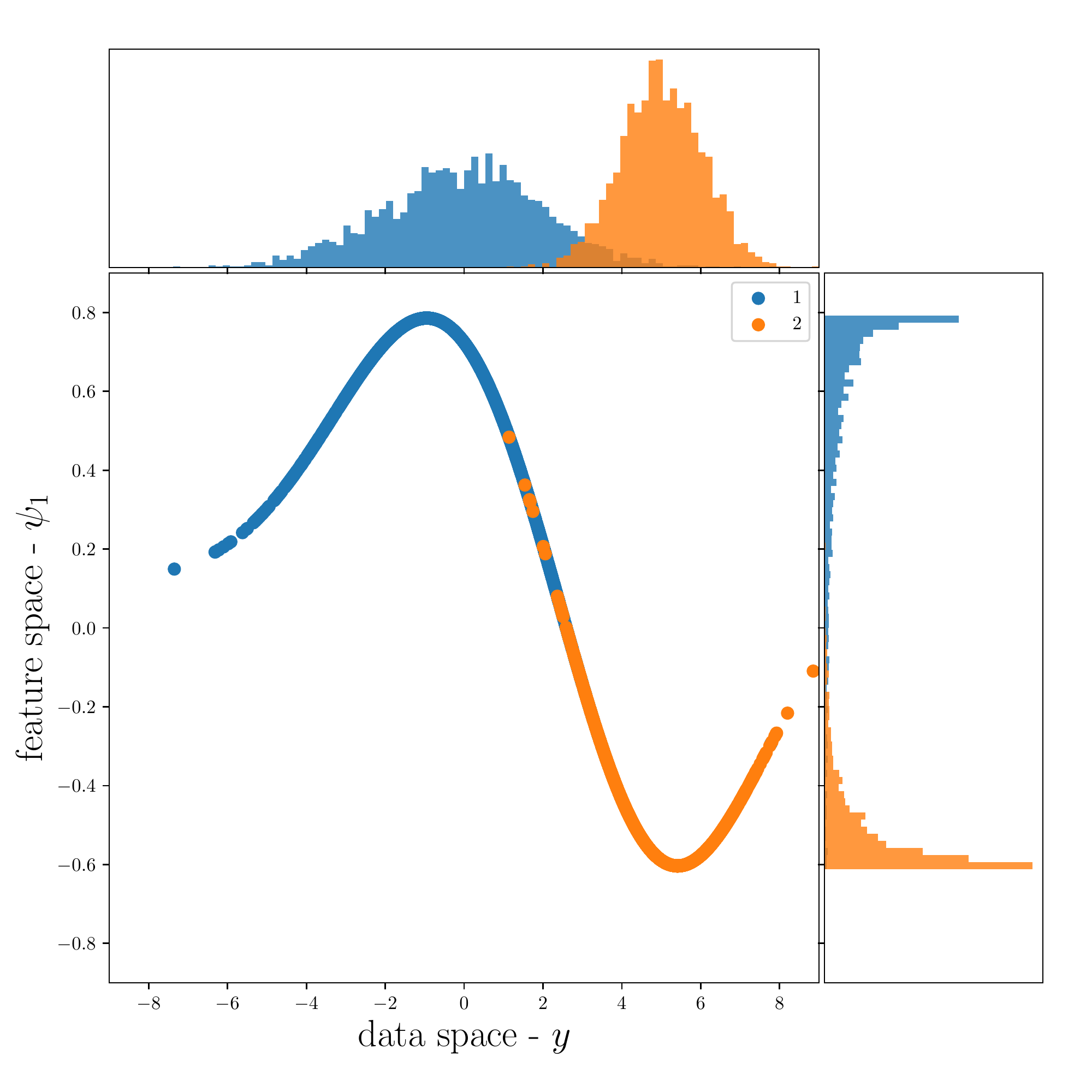}
     \caption{Mapping from data space to the KPCA feature space $\boldsymbol{\psi}_N$. Starting from the top distribution in data space, points are transformed through the learned function in the middle into the first component of the feature space, shown on the right. Samples from two underlying distributions are denoted with different colors for better visualization.}
     \label{fig:d_to_f}
\end{figure}

For the example above, one can show the first component of the feature space $\psi_1(y)$ (see Figure \ref{fig:d_to_f}).  Samples from the modes in the distribution are distinguished by different colors, for better visualization.  Starting from the distribution on the top, we pass it through the (learned) transformation shown in the middle, getting the distribution on the right. As expected, the two modes (blue vs orange) are much better separated in the first component of the feature space, $\psi_1$, than they were in data space, $y$.

Contrary to linear PCA however, an exact inverse transform to return from $\psi_1$ back to data space generally does not exist. Therefore, we define the inverse transform using kernel ridge regression. In  linear ridge regression from $\boldsymbol{\psi}_N$ back to $y$, we would minimize the mean square error over the data:
\begin{equation}
    \sum_i \left(y_i - \mathbf{w}^T \boldsymbol{\psi}_N\right)^2 + \frac{\lambda}{2} ||\mathbf{w}||^2 \, ,
\end{equation}
where $\mathbf{w}$ are the weights and second term is a standard $l_2$ regularization. 
However, as the inverse function is highly non-linear, we firstly transform $\boldsymbol{\psi}_N$ into another (possibly infinite) feature space $\phi$, and learn the transformation $\widetilde{\mathbf{w}}$ back to $y$. The minimization is then:
\begin{equation}
    \sum_i \left(y_i - \widetilde{\mathbf{w}}^T \phi\left(\boldsymbol{\psi}_N\right)\right)^2 + \frac{\lambda}{2} ||\widetilde{\mathbf{w}}||^2 \, .
\end{equation}
One can prove that the feature space $\phi$ does not have to be accessed and is only implicitly defined by the kernel $\widetilde{\kappa} \left((\boldsymbol{\psi}_N)_i, (\boldsymbol{\psi}_N)_j\right) = \phi\left((\boldsymbol{\psi}_N)_i\right) \cdot \phi\left((\boldsymbol{\psi}_N)_j\right)$. 

\begin{figure}
     \centering
     \includegraphics[width=0.9\linewidth]{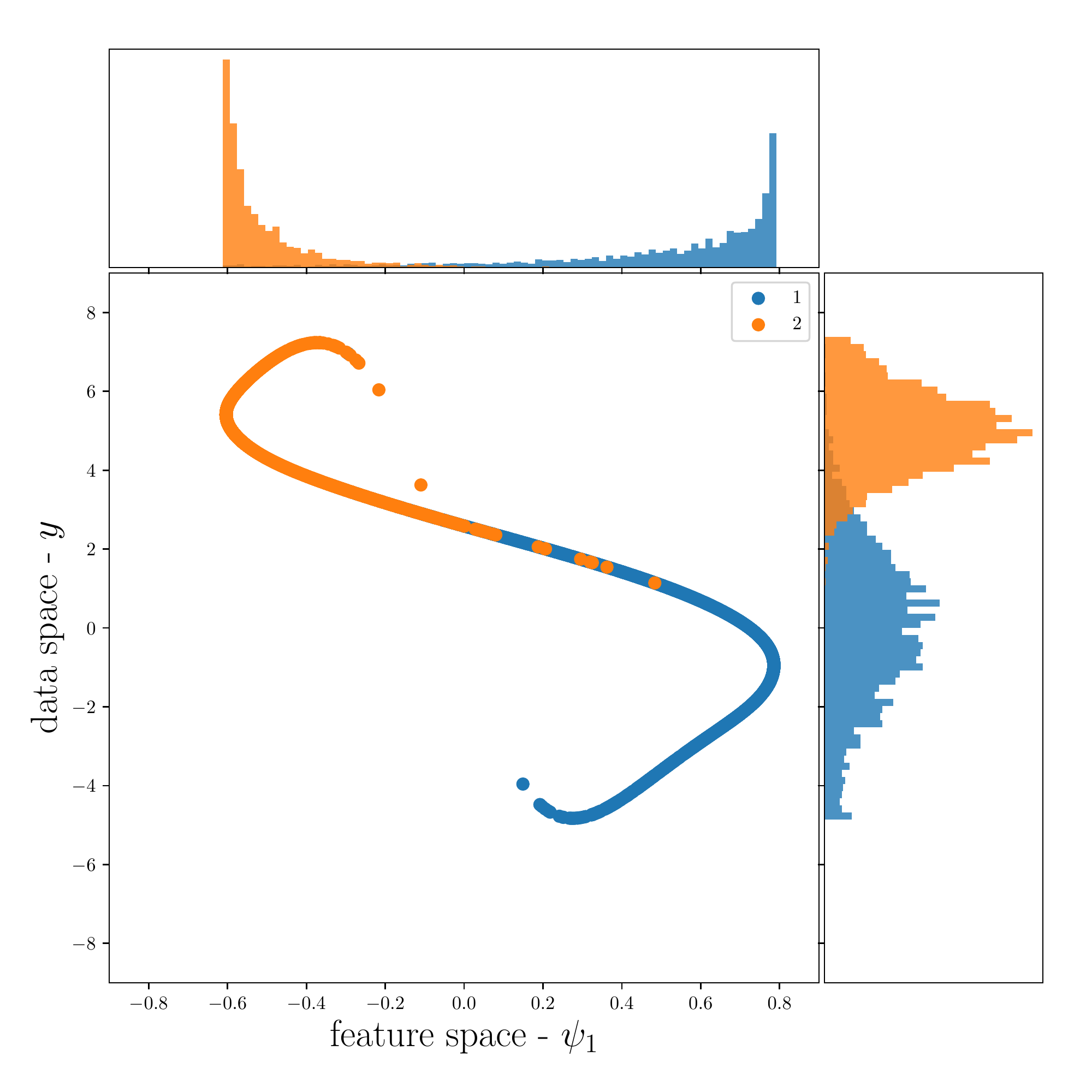}
     \caption{Mapping from the feature space $\boldsymbol{\psi}_N = (\psi_1, \psi_2, ...)$ back to the data space, $y$. Kernel ridge regression feature space $\phi$ is never accessed and only defined by the kernel $\widetilde{\kappa}$. Samples from the two underlying modes are separated in color for better visualization.}
     \label{fig:f_to_d}
\end{figure}

In Figure \ref{fig:f_to_d}, we show the results of such procedure. We can see that the learned mapping is indeed non-linear and the initial distribution is well preserved.

\bsp	
\label{lastpage}
\end{document}